\documentclass[a4paper,12pt]{article}

\usepackage[margin=2.5cm]{geometry}

\usepackage{textgreek}

\usepackage{titlesec}
\usepackage{setspace}
\usepackage[all]{nowidow}
\usepackage{amsmath}
\usepackage[version=4]{mhchem}
\usepackage{siunitx}
\usepackage{float}
\usepackage{graphicx}
\usepackage{rotating}
\usepackage{caption}
\usepackage{subcaption}
\usepackage[
backend=biber,
style=vancouver
]{biblatex}

\renewbibmacro*{volume+number+eid}{%
    \printfield{volume}%
    \printfield{number}%
}\DeclareFieldFormat[article]{number}{\mkbibparens{#1}}

\DeclareFieldFormat*{pages}{\addspace\mkcomprange{#1}}

\DeclareFieldFormat[article,periodical]{volume}{\mkbibbold{#1}}

\DeclareSIUnit \nucleon {u}
\DeclareSIUnit \ion {ion}
\DeclareSIUnit \frame {frame}
\DeclareSIUnit \Hit {Hit}
\DeclareSIUnit \vertices {vertices}
\DeclareSIUnit \vertex {vertex}

\usepackage{hyperref} % Load last

\graphicspath{{.}}

\newcommand{\cion}{% Prevent leading space
    \ce{^12C^{6+}}% Prevent trailing space
}

\newcommand{\singleFig}[5][!htbp]{%
    \begin{figure}[#1]%
        \centering%
        \captionsetup{width=15cm}%
        \includegraphics[width=15cm]{#2}%
        \caption[#5]{#4}%
        \label{fig:#3}%
    \end{figure}%
}

\newcommand{\rotFig}[4]{%
    \begin{sidewaysfigure}%
        \centering%
        \captionsetup{width=15cm}%
        \includegraphics[width=247mm]{#1}%
        \caption[#4]{#3}%
        \label{fig:#2}%
    \end{sidewaysfigure}%
}

\newcommand{\doubleFig}[6][!htbp]{%
    \begin{figure}[#1]%
        \centering%
        \captionsetup{width=16cm}%
        \begin{subfigure}[t]{7.9cm}%
            \centering%
            \includegraphics[width=\linewidth]{#2}%
            \phantomsubcaption%
            \label{fig:#4_a}%
        \end{subfigure}%
        \hspace{0.2cm}%
        \begin{subfigure}[t]{7.9cm}%
            \centering%
            \includegraphics[width=\linewidth]{#3}%
            \phantomsubcaption%
            \label{fig:#4_b}%
        \end{subfigure}%
        \caption[#6]{#5}%
        \label{fig:#4}%
    \end{figure}%
}

\newcommand{\figref}[1]{% Prevent leading space
    Figure \ref{fig:#1}% Prevent trailing space
}

\begin{document}
    \begin{center}
        \Large{A parallel, pipeline-based online analysis system for Interaction Vertex Imaging}
    \end{center}
    \begin{center}
        \textbf{Devin Hymers}\textsuperscript{1},
        \textbf{Sebastian Schroeder}\textsuperscript{1},
        \textbf{Olga Bertini}\textsuperscript{2},
        \textbf{Johann Heuser}\textsuperscript{2},
        \textbf{Joerg Lehnert}\textsuperscript{2},
        \textbf{Christian Joachim Schmidt}\textsuperscript{2},
        \textbf{and}
        \textbf{Dennis M\"ucher}\textsuperscript{1}
    \end{center}
    
    \textsuperscript{1}Institute for Nuclear Physics, University of Cologne, 50937 Cologne, Germany
    
    \textsuperscript{2}GSI GmbH, 64291 Darmstadt, Germany
    
    \section{Abstract}
    \label{sec:online_abstract}
    
    \textbf{Objective}\\
    Interaction vertex imaging (IVI) is used for range monitoring in carbon ion radiotherapy, detecting depth differences between Bragg peak positions. Online range monitoring, which provides feedback during beam delivery, is particularly desirable, creating an opportunity to detect range errors before the treatment fraction is completed. Incorporating online range monitoring into clinical workflows may therefore improve the safety and consistency of radiotherapy.
    
    \textbf{Approach}\\
    The data analysis system was broken into a task-parallel pipeline approach, to allow multiple analysis stages to occur concurrently, beginning during acquisition. Computationally-expensive operations were further parallelized to reduce bottleneck effects. Data collected from irradiation of homogeneous plastic phantoms was replayed at the same rate it was initially acquired, to mimic data acquisition, and the time required to determine a range shift was measured.
    
    \textbf{Main Results}\\
    With an optimized pipeline, the delay between the end of irradiation and the determination of a range shift is consistently less than \qty{200}{\milli\s}. The majority of this time is associated with the final range shift determination, with a minor effect from the time required to analyze the last data packet. The most significant contribution to an optimized analysis workflow is the formation of clusters, requiring almost \qty{50}{\percent} of compute time.
    
    \textbf{Significance}\\
    This system is the first IVI implementation to achieve clinically-relevant online analysis times. The \qty{200}{\milli\s} time required to determine a range shift is less than the time required to accelerate a new spill in a synchrotron, and is comparable to the time required for reacceleration if multiple energies are delivered in the same spill. Clinical implementation of online range monitoring would allow treatment to be quickly paused or aborted if significant range errors are detected.
    
    \section{Introduction} 
    \label{sec:online_intro}
    
    A growing modality for external-beam radiation therapy is Carbon Ion Radiotherapy (CIRT), which uses a \cion{} beam to deliver dose to the tumour \autocite{malouff_carbon_2020}. Unlike the more common photon-based radiotherapy, the charged beam used in CIRT loses energy quasi-continuously in the patient, resulting in a beam which slows and stops within the patient. The dose distribution resulting from this behaviour is inverted relative to that of a photon beam, resulting in the dose maximum, or Bragg peak (BP) being located where the beam stops \autocite{amaldi_radiotherapy_2005}. Therefore, CIRT is capable of delivering highly precise, conformal dose distributions, maintaining a high dose to the tumour while sparing the surrounding healthy tissue \autocite{amaldi_radiotherapy_2005}. As most tumours are significantly larger than can be uniformly irradiated by a single BP, CIRT treatments commonly involve a constellation of BP positions and intensities \autocite{kramer_treatment_2000}. Full tumour coverage is achieved by varying the lateral position through magnetic steering of the beam, and varying the BP depth by changing beam energy \autocite{haberer_magnetic_1993}. As with most forms of external beam radiation therapy, treatments are also commonly fractionated. The total prescribed dose is delivered in daily irradiations over a period of several weeks, allowing cellular repair mechanisms in healthy tissue to moderate the damage to non-cancerous cells \autocite{amaldi_radiotherapy_2005}.
    
    However, the dosimetric advantage of CIRT also introduces the additional challenge of monitoring the Bragg peak position, to ensure it is located directly on the tumour. Currently, safety margins on the range of millimeters are required around a tumour, to account for uncertainties in beam delivery and stopping range \autocite{kelleter_-vivo_2024, andreo_clinical_2009}. Range monitoring (RM) methods are used to measure the BP depth, with the goal of improving the consistency of dose delivery, and reducing the required safety margins \autocite{finck_study_2017, hymers_intra-_2021}. Achieving these goals would allow a reduction in the dose delivered to healthy tissue surrounding the tumour, while maintaining the currently achievable level of tumour control.
    
    One RM method which is currently undergoing clinical trials for use in CIRT is Interaction Vertex Imaging (IVI) \autocite{fischetti_inter-fractional_2020, kelleter_-vivo_2024}, which uses external detectors to monitor prompt secondary charged particles (primarily protons) created by beam-patient reactions \autocite{amaldi_advanced_2010}. These external detectors track the particle, and backproject its trajectory into the patient to approximate the location of the reaction (or interaction vertex) which produced it \autocite{finck_study_2017,  henriquet_interaction_2012, hymers_intra-_2021}.
    As the secondary protons experience the same sort of quasi-continuous energy loss in the patient as the primary \cion{} beam, the majority of the reconstructed interaction vertices are superficial to the BP position \autocite{henriquet_interaction_2012, gwosch_non-invasive_2013, hymers_monte_2019}. Because of this limitation, IVI is best suited for relative RM, evaluating the consistency of BP positioning between fractions delivered on different days, or between BPs delivered at different depths in the same fraction \autocite{fischetti_inter-fractional_2020, kelleter_-vivo_2024}. These methods show promise for allowing RM of each fraction, and have been shown to be sensitive to BP range differences of millimeters to hundreds of micrometers \autocite{finck_study_2017, hymers_intra-_2021}. Such range differences may stem from patient movement, differences in patient positioning for each fraction delivery, or variations in patient internal structure during treatment \autocite{ammazzalorso_dosimetric_2014, handrack_sensitivity_2017}.
    
    The ideal implementation of IVI would provide online RM \autocite{muraro_monitoring_2016, parodi_vivo_2018}, allowing the consistency between fractions to be applied as an additional safety check in beam delivery. If online IVI were to detect an unallowably-large deviation in BP range in the early phase of a fraction, the remainder of the fraction could be paused, and either modified or aborted to avoid overirradiation of healthy tissue, or underirradiation of the tumour itself. Such a deviation in expected BP range could also trigger a repeat of the anatomical imaging which is performed during treatment planning, to visualize any changes in patient internal structures which would otherwise have remained undetected \autocite{fischetti_inter-fractional_2020, kelleter_-vivo_2024}. Therefore, the successful implementation of online RM would potentially lead to not only an improvement in treatment safety from reduced irradiation of healthy tissue, but also in treatment efficacy, by reducing the probability of incompletely irradiating the target volume in any treatment fraction \autocite{fischetti_inter-fractional_2020}.
    
    To achieve online RM, a fast data acquisition and analysis system is required \autocite{hymers_intra-_2021, ghesquiere-dierickx_detecting_2022}. IVI is well-suited for online RM, as all data collection occurs during the beam-on period of irradiation, and the typical latency between a primary ion entering the patient and the associated secondary particles being detected is on the order of nanoseconds. The leading contribution to this latency is the physical travel time of the particles, which despite moving at a significant fraction of $c$, must cover distances on the order of tens of centimeters \autocite{piersanti_measurement_2014}. Therefore, the data is made available for processing near-instantaneously, and the only limitation is the speed with which analysis can be performed. More rapid analysis is clearly advantageous, allowing earlier detection and abortion of an incorrectly-ranged fraction. However, there are characteristic timescales for which these advantages become more significant.
    
    Clinical \cion{} beams are currently delivered by synchrotron accelerators, such as the synchrotron at the Heidelberg Ion-Beam Therapy Center (HIT). The properties of a synchrotron naturally form a pulsed beam macrostructure due to the pause between spills as the accelerator is refilled and primary particles are reaccelerated. This refilling is often combined with a change in primary beam energy, to irradiate a different depth in the patient. At HIT, typical spills last between \qtyrange[range-phrase=~and~]{1.5}{5.0}{\s}, with a gap of approximately \qty{3.0}{\s} for particle injection and acceleration between spills \autocite{schoemers_christian_beam_2023, schoemers_first_2017}. These spills may also be delivered with gaps of up to \qty{1.0}{\s}, for irradiation of disjoint regions on the same isoenergy surface in the patient \autocite{hoffmann_beam_2008}. Spill gaps at HIT may also include a reacceleration phase of \qtyrange{100}{500}{\milli\s}, during which the beam energy can be increased, allowing irradiation of multiple different isoenergy slices within the same spill. This reacceleration phase reduces the time required to deliver an entire treatment, improving efficiency and reducing the amount of time the patient is required to remain in the treatment room \autocite{schoemers_first_2017}. Therefore, there are two relevant targets for analysis speed: less than \qty{3.0}{\s}, for processing of a complete spill of duration \qty{5.0}{\s}; and on the order of \qty{100}{\milli\s}, for processing of a partial spill of order \qty{1.0}{\s}, with or without reacceleration. Meeting either of these benchmarks corresponds to a timescale in which it is possible to detect a range error before the next spill or partial spill is delivered. Once an error is detected, irradiation may be aborted within \qty{200}{\micro\s}, maximizing the potential for dose sparing \autocite{peters_spill_2008}.
    
    This work provides an in-depth description of a system for performing online IVI, as well as a discussion of factors affecting its rate performance. This system is validated to consistently meet clinically-relevant targets for data analysis speed, demonstrating the ability generate an abort command on the same timescale as that required to change the beam energy.
    
    \section{Data Acquisition Hardware}
    \label{sec:online_hardware}
    
    Data collection for RM was performed using the fIVI Range Monitoring System, previously described in a separate publication \autocite{hymers_evaluation_2025}.
    
    \subsection{Physical Configuration}
    \label{sec:online_tracker}
    
    The tracker of the prototype fIVI Range Monitoring System was designed and manufactured at the University of Cologne, using sensors and readout electronics developed by GSI \autocite{heuser_technical_2013}. The resultant device consists of two layers of \qty{300}{\micro\m} thick position-sensitive silicon detectors, the front layer with a sensitive area of \qtyproduct[product-units=power]{6.0x6.0}{\cm}, and the rear layer with a sensitive area of \qtyproduct[product-units=power]{6.0x12.0}{\cm}. These two layers are placed \qty{12.0}{\cm} apart inside a light-tight aluminum enclosure, with the front layer \qty{3.5}{\cm} from the \qty{16}{\micro\m} aluminum entrance window. The sensors are positioned such that their center points are at the same height above the bottom of the enclosure.
    
    The position-sensitivity of these sensors was achieved through double-sided strip-segmentation. However, rather than the conventional orthogonal strips on opposite planar faces, these sensors had a stereo angle of only \ang{7.5}, with axially-oriented strips on the n side of the silicon, and angled strips on the p side. This choice of segmentation allowed readout of the entire sensor along a single \qty{60}{\mm} edge, at the cost of reducing the total number of unique intersections between segments on opposite faces, and consequently reducing the spatial resolution of the sensor in an axis-dependent fashion.
    
    \subsection{Readout Electronics}
    \label{sec:online_electronics}
        
    Each sensor is read out using the SMX 2.2 application-specific integrated circuit (ASIC) \autocite{kasinski_sts-xyter_2014, kasinski_back-end_2016}. Each ASIC directly digitizes interactions in 128 segments from a single side of the sensor, connected via analog microcables. Readout of all \num{1024} segments from a single side requires eight ASICs, which are collected on a single printed circuit board, termed the `Front-End Board' (FEB). The SMX 2.2 ASIC provides a fast and compact readout system, with \qty{6.25}{\nano\s} timing precision in a 14-bit timestamp stemming from a \qty{160}{\mega\hertz} timing clock. High-precision timing is achieved by latching the timestamp through a separate circuit using a fast shaper. Energy sensitivity is implemented through pulse height analysis; to achieve fast timing performance, shaping times on the order of \qty{100}{\nano\s} were used. The height of the output pulse is converted to a 5-bit digital value through a sequence of 31 discriminators. This relatively low energy resolution is an acceptable compromise to achieve fast timing and reset performance. The dynamic range of the analog-to-digital converter is \qty{100}{\femto\coulomb}.
    
    Each SMX communicates digitally with upstream data processing electronics, implementing a custom communication protocol with a constant 24-bit frame size and 8b/10b encoding on one or two low-voltage differential pairs per ASIC. Although multiple SMXs are mounted to the same FEB, and share a common downlink for control signalling, each ASIC establishes its own upstream link(s) with the data processing board. Signalling over these links uses an \qty{80}{\mega\hertz} clock, which is doubled to provide the timing clock used in assigning timestamps to trigger events \autocite{kasinski_protocol_2016}.
        
    Each GBTX Emulator (GBTxEMU) data processing board aggregates data from two FEBs, for a total of up to 28 uplinks. These boards communicate via optical link with the data acquisition PC over the bidirectional gigabit transciever (GBTX) protocol, and are responsible for both relaying acquired data from the FEBs to the PC, and relaying commands in the opposite direction. Data received by the acquisition PC is placed directly into system memory through direct memory access (DMA) from a PCI Express add-in card, the GBTxEMU Readout Interface (GERI), which also provides a common control interface \autocite{zabolotny_versatile_2023}.
    
    The two sensors of the fIVI Range Monitoring System are read out by \num{32} SMX ASICs, each with two physical elink uplinks for a total of \num{64}. A full capacity readout system would therefore require the bandwidth of three GBTxEMU boards; however, due to hardware limitations of the current setup which require each FEB be connected to only one GBTxEMU board, such a system can only be achieved with one readout board per FEB. This approach results in significant wasted bandwidth in the acquisition system. Another approach, which maximizes the utilization of each GBTxEMU board, divides the \num{28} elinks between two FEBs, allocating \num{16} elinks to one FEB, and the remaining \num{12} to another. Each ASIC on the FEB allocated \num{16} elinks is then able to take data at its full bandwidth of two uplinks per ASIC, while each ASIC on the FEB allocated \num{12} elinks is only guaranteed access to a single uplink, although four of the eight do benefit from a second uplink. The latter approach is used in the fIVI Range Monitoring System, as it allows all four FEBs to be connected using only two GBTxEMU boards, with no unused elinks, maximizing the space and power efficiency of the system. As the front sensor is closer to the target and covers a larger solid angle, it is therefore expected to experience a significantly higher event rate than the rear sensor in all situations. Therefore, the maximum bandwidth is allocated to the front sensor, using the topology shown in \figref{daq}.
    
    \singleFig{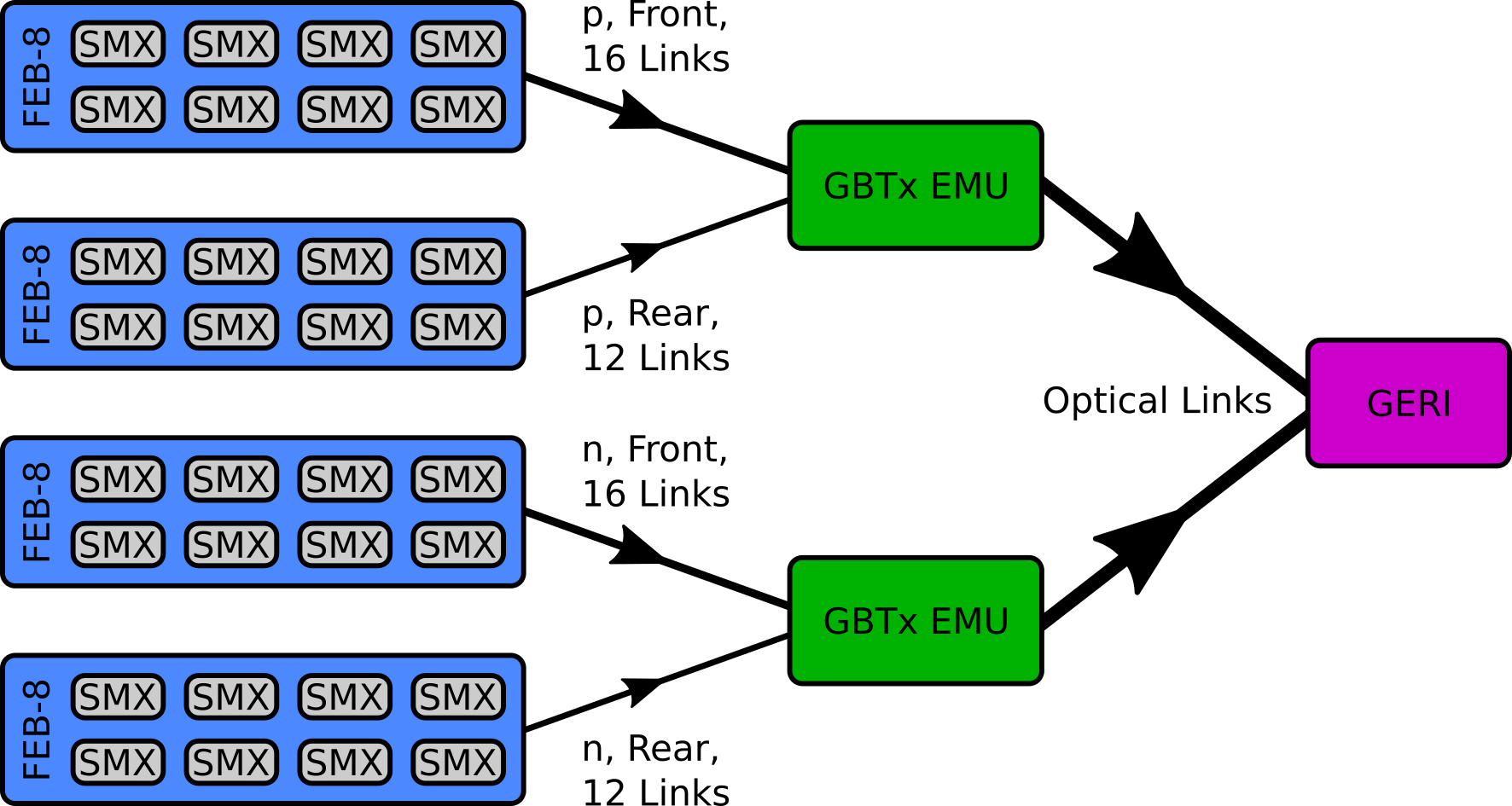}{daq}{Schematic of data flow within the data acquisition system. Each GBTxEMU readout board (green) aggregates data from two front-end boards (blue), which are each responsible for reading out an entire sensor side. Aggregated data from the GBTxEMU is sent along an optical link to the GERI readout interface (magenta), which combines the entire system into a single datastream.}{Schematic of data flow within the data acquisition system.}
    
    Both GBTxEMU boards are then connected to a single GERI, installed in a Dell Optiplex Tower Plus  workstation. This workstation is equipped with \qty{64}{\giga\byte} of DDR5 system memory, providing sufficient DMA space for the GERI driver while not compromising the working memory needed for data processing and analysis. The Intel i9 13900K CPU with \num{24} physical cores and \num{32} logical threads allow sufficient processing power for running the control and analysis software on a single machine. The heterogeneous architecture of this processor combines excellent single-thread performance on eight performance cores operating at up to \qty{5.80}{\giga\hertz}, with good support for highly parallel workloads through the inclusion of 16 efficiency cores.
    
    \section{Data Acquisition Software}
    \label{sec:online_software}
    
    \subsection{Control Software}
    \label{sec:online_control}
    
    The control interface for the fIVI Range Monitoring System is written in Python 3, and is based on Python libraries which model the GBTxEMU and GERI hardware and firmware, as well as implementing the GBTX and HCTSP (Hit and Control Transfer Synchronous Protocol) communication protocols \autocite{kasinski_protocol_2016}. On top of these libraries, a model of the SMX ASIC has been implemented, modelling the internal state of all control registers. The primary control software can run in either calibration mode, using the internal pulse generator on a specific SMX to calibrate the analog-to-digital conversion circuitry for the current ASIC configuration, or in data acquisition mode, using broadcast commands to synchronize, start, and stop data recording.
    
    The HCTSP communication protocol employed by the SMX uses a constant 24-bit uplink frame size, with several different frame types defined. The most important of these are the Hit frame, which corresponds to a self-trigger event of a single channel, and the TS\_MSB frame, which contains the six most significant bits of the SMX's 14-bit timestamp. By transmitting these six bits separately from the Hit frame, lossless data compression is possible at high event rates or event bursts, in which multiple adjacent channels are triggered simultaneously. To prevent ambiguity in timing which may otherwise occur at low data rates, when no trigger events are available to send, automatically-generated TS\_MSB frames are also sent every time the five most significant timestamp bits are updated. Using these automatically-generated TS\_MSB frames, later portions of the readout and data processing system are kept aware of the passage of time, allowing the reconstruction of a global timestamp for each Hit frame. When more than one elink is active on a given SMX, Hit frames are sent over only one link. TS\_MSB frames, however, are generated and sent for each link independently, such that the same automatically-generated TS\_MSB frames appear on all elinks, if no Hit frames are available.
    
    In data acquisition mode, the startup procedure first establishes connection with the GERI, then synchronizes GBT links with each GBTxEMU board, before finally synchronizing elinks with each SMX ASIC. The full synchronization procedure is carried out on each execution; the quick synchronization procedure used in other setups with the same readout hardware is not yet implemented for the fIVI Range Monitoring System \autocite{kasinski_protocol_2016}. Once communication has been confirmed with all ASICs, the ASIC configuration and previously-determined calibration settings are loaded from file into the software SMX model, then written to the corresponding hardware. The entire setup process takes approximately two minutes, the majority of which is dedicated to the elink synchronization process. DMA access through the GERI is also configured in packeted mode, finalizing one packet every \qty{100}{\milli\s}, containing all the data received since the previous packet was delivered.
    
    Multiple discrete datasets may then be acquired on a single run of the control software, without repeating the setup. At the start of each dataset, a synchronization procedure is implemented, using commands broadcast to the entire system. First, the SMX Hit frame outputs are disabled, preventing interactions recorded by the analog to digital converters from being queued for output over the elinks. Then, a start command is given to the DMA data acquisition, causing the GERI to begin listening for data to pass via DMA. This order of operations ensures that no data is included in the run before the synchronization procedure has completed, but that the synchronization motif is still visible in the GERI DMA datastream. Synchronization is achieved by loading a timestamp of \num{0} into each SMX, then simultaneously applying the loaded value across all ASICs; the sequence of ACK frames sent in response to the load and apply commands, along with the automatically-generated TS\_MSB frames, provides a synchronization motif from each ASIC, which sets the time zero point for this elink. Finally, the SMX Hit frame outputs are re-enabled, allowing synchronized data to flow through the entire readout chain. Ending the acquisition is comparatively simple, achieved by stopping the GERI DMA engine.
    
    Communication with the GERI DMA engine during data acquisition consists of two parts. The first is a direct communication with the GERI driver, writing registers which reset the DMA engine at acquisition start to prepare for writing a new data block, and enabling or disabling DMA. The second is a call via the readout software library, to prepare for processing a new file. As the readout software is written in C, these calls are managed using the ctypes Python library, which creates Python bindings for compiled C libraries.
    
    \subsection{Readout Software}
    \label{sec:online_readout}
    
    The readout software is responsible for low-level communication with the GERI data acquisition interface directly through the driver. This library is written in C, and is heavily based on the example processing code distributed with the GERI driver \autocite{zabolotny_versatile_2023}. In addition to the primary function of accepting data placed in the DMA space by the GERI and forwarding that data for processing, the readout software is responsible for setup of parameters related to DMA access and data delivery, such as initializing the DMA buffer space, configuring its size, and providing that information to the GERI hardware at the start of each acquisition. After all acquisitions are complete, this software is also responsible for cleaning up the DMA space.
    
    During acquisition, the readout software waits for data to be available in the DMA space. Once notified of a completed packet delivered into the DMA space by the GERI, the entire block of data is written to the output stream, which may either be directly to file, for offline analysis, or to a named pipe, which bypasses the slow write to disk and sends the data directly to an online analysis system. Using the UNIX construct of named pipes allows a consistent interface for both the readout and analysis software, regardless of whether analysis is completed online or offline, as the pipe may be utilized in exactly the same way as a real on-disk file. After the packet has been written, receipt of the packet is confirmed to the driver, informing the GERI that the DMA space containing that packet is available to be overwritten with future data.
    
    As the processing completed directly by the readout software is minimal, data is processed in single-packet mode \autocite{zabolotny_versatile_2023}. This mode eliminates the need for synchronization overhead between multiple threads, as the analysis software is designed to accept an ordered, serial datastream, and data can be buffered outside of the DMA space, if necessary. It is most expedient to clear the DMA space as quickly as possible, then perform parallelized analysis of each packet, as the timestamp associated with any interaction in a given packet can only be perfectly determined after processing all preceding packets.
    
    \section{Analysis Software}
    \label{sec:online_analysis}
    
    The analysis software is written in C++, and has been developed from scratch to suit the needs of the fIVI Range Monitoring System. The role of this software is to process the raw data stream delivered from the GERI: recombine Hit and TS\_MSB frames into individual channel trigger events with full timing information; form clusters of adjacent and coincident trigger events (which likely correspond to interactions of the same particle); and match coincident clusters from opposite sides of the same sensor to position those particle interactions in the lab frame. For range monitoring using fIVI, coincident hits from both detector layers are combined into particle tracks, which are then reconstructed using fIVI to produce a vertex distribution, and used to determine a range shift as compared to a previous measurement. All of these coincidence windows are independently configurable.
    
    To improve reconstruction performance, this analysis is performed in a parallel, pipelined manner. Each stage of analysis occurs in one or more separate threads, with additional threads being created for object-level parallelism wherever synchronization between objects is not required, as shown in \figref{pipeline}. For instance, reconstruction of hits on two different sensors occurs in two separate threads, and both data streams are combined to feed into the next stage of analysis. All communication between threads is unidirectional, and uses a single-producer, single-consumer lockfree queue structure (SPSC queue), which significantly reduces the overhead of synchronization in a highly parallel environment. This data structure is tuned to the 64-byte cache line size of the x86 processor family (including the i9 13900K CPU on which this software was tested), and employs relaxed memory ordering to reduce the performance penalties stemming from contention and cache invalidation, when many small objects are being moved through the queue.
    
    \singleFig[t]{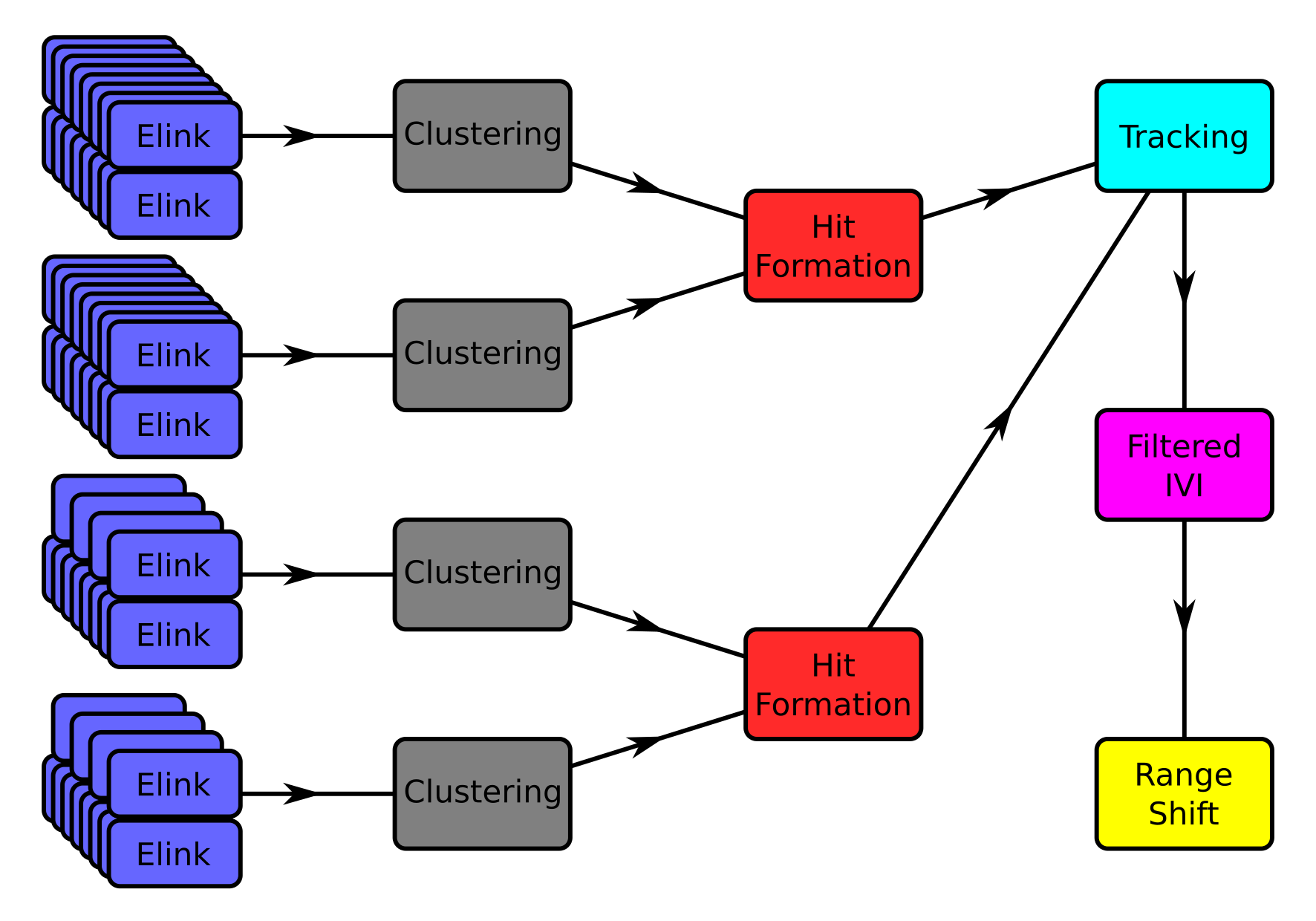}{pipeline}{Schematic of the data analysis pipeline. Each uplink is processed in its own thread (blue). All links corresponding to the same side of a sensor are combined at the clustering phase (grey), where groups of adjacent and coincident trigger events are formed. Coincident clusters from opposite sides of the same sensor form hits (red), and coincident hits from different layers of the tracker are used to form tracks (cyan). Each track is then independently evaluated using the filtered IVI method (magenta); the resultant interaction vertices are used in the range shift determination process (yellow). Each element represents a single thread of execution; the counts are reflective of the evaluated configuration. However, the system is highly flexible, and each element (with the exception of hit formation) is able to accept an arbitrary number of inputs.}{Schematic of the data analysis pipeline.}
    
    Due to the high level of parallelism, more threads are created than there are physical or logical CPU cores on the analysis system. To reduce contention over critical resources, whenever a thread waits for input data from an empty SPSC queue, or waits to pass data along to a full SPSC queue, the thread sleeps briefly, which allows it to be descheduled and yield its time to another thread. In this way, threads are primarily active while they are performing useful work, and only briefly active while waiting for more data to process. Using a model in which threads may be descheduled, rather than blocking and waiting for notifications, significantly reduces the overhead of inter-thread communication, which is important for a pipeline approach where many small objects are frequently passed between threads.
    
    \subsection{Trigger Event Formation}
    \label{sec:online_trigger}
    
    As each elink generates its own timing signals through the TS\_MSB frames, the reconstruction of a complete timestamp for each trigger event must take place on a per-elink basis. Therefore, the incoming data stream is split between threads, using the 8-bit tag assigned by the GERI, so that each thread may process the stream from a single elink without interference.
    
    The goal of elink processing is to supplement each Hit frame with a full timestamp, in place of the least significant bits reported in the frame. This timestamp may be constructed by combining the least significant bits from the Hit frame with the bits from the most recent TS\_MSB frame. Validation is possible using the two bits of overlap included in both frames. However, this is not sufficient data to generate a full timestamp, as the \num{14} total bits of timestamp kept by the SMX represents a duration of only \qty{102.4}{\micro\s}, as the timer increments on every rising and falling edge of the \qty{80}{\mega\Hz} elink clock. To keep absolute time on the seconds-long timescale of a synchrotron spill, or minutes of an entire treatment, it is necessary to also keep track of the number of times this timestamp rolls over.
    
    For the purpose of tracking timestamp rollovers, automatically-generated TS\_MSB frames are sent every \qty{3.2}{\micro\s}, or every time the ninth bit of the timestamp changes, so long as this frame does not interfere with sending actual Hit data. Although these frames are identical to the TS\_MSB frames which are associated with Hit frames, they can be effectively differentiated by their relative position in the datastream: the automatically-generated TS\_MSB frames should never come directly before a Hit frame, as shown in \figref{sequencing}. Furthermore, the automatically-generated TS\_MSB frames report a different set of bits from the TS\_MSB frames associated with Hits: bits \verb|<12:7>|, rather than bits \verb|<13:8>|. Therefore, only in rare situations is confusion possible, such as when an entire frame is lost or missed, and these cases can be noticed and filtered out, as the overlap bits will not match as expected. The bitshift between the two classes of TS\_MSB frame provides a second advantage as well: it provides one bit of overlap between the number of times the automatically-generated TS\_MSB frames have experienced a rollover and the top bit of the Hit-associated TS\_MSB frame, which can be used as an additional consistency check. In this way, the duration of data acquisition may be extended well beyond the time required for treatment through the construction of a 64-bit timestamp. The combination of a Hit frame and the extended 64-bit timestamp, termed a single-channel trigger event, is passed to the next stage of reconstruction, while all other frame types are noted and discarded.
    
    \singleFig[t]{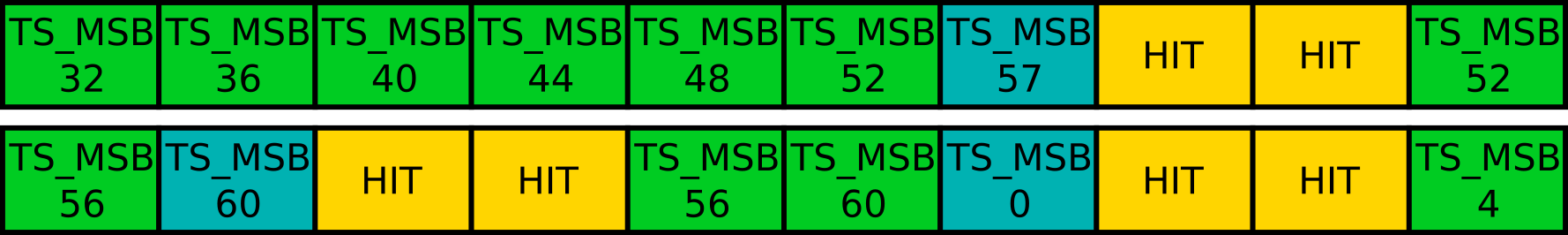}{sequencing}{Representative example of the sequence of data in a single elink. TS\_MSB frames (green) are automatically generated at regular intervals to record the passage of time. When Hit frames (yellow) are ready to be sent, Hit-associated TS\_MSB frames (blue) are also generated if necessary. These Hit-associated frames are only required if the most recent TS\_MSB frame sent over the link does not match the most significant bits of the upcoming Hit. Three cases are represented in this figure. In the upper right, the Hit-associated TS\_MSB frame is clearly not a part of the automatically-generated sequence. In the lower left, the Hit-associated TS\_MSB frame could be a part of the automatically-generated sequence, but it is clearly distinguished by being followed by Hit frames. Once the Hit frames have been sent, the automatically-generated timestamp is reasserted over the link. In the lower right, no Hit-associated TS\_MSB frame is generated, as the automatically-generated TS\_MSB frame is already valid for these Hits. In this case, the automatically-generated TS\_MSB frame is treated as a Hit-associated frame, so no additional TS\_MSB frame is required, and the passage of time resumes with the next automatically-generated frame.}{Representative example of the sequence of data in a single elink.}
    
    The one exception to the strict separation of elinks during this stage of analysis is in the synchronization motif which occurs at the start of each run. Like Hit frames, the ACK frames sent in response to the synchronization commands are delivered on only a single uplink per SMX ASIC, while most ASICs in this system use multiple uplinks. Therefore, when an ACK frame is detected by the analysis software, that frame is duplicated along all uplink data streams which correspond to the SMX, so that the synchronization motif is recognizable in all cases.
    
    As the sensors are quite large, and there is a strong spatial dependence of the count rate produced during IVI measurements, different regions of the sensor may experience significantly different event rates \autocite{ghesquiere-dierickx_investigation_2021, hymers_intra-_2021}. These rate differences can extend to individual ASICs, which read out defined segments of the sensor area. To prevent deadlocks in the analysis system which may occur when searching for coincidences between a high-rate region and a low-rate region, the analysis software periodically sends special `keepalive' events, which serve a similar purpose to the automatically-generated TS\_MSB frames. These keepalive events do not participate in analysis, but provide later stages with a concept of the passage of time, indicating that no future events will ever arrive from this source with an earlier timestamp than the keepalive.
    
    \subsection{Cluster Formation}
    \label{sec:online_cluster}
    
    The cluster formation process accepts trigger events from all elinks which contribute to a given side (p or n) of a single sensor. These events are sorted based on their timestamp, and assigned to the appropriate segment of the sensor. The primary clustering algorithm examines all segments, and selects the one with the earliest timestamp to initiate cluster formation. All continuous and coincident trigger events are added to the cluster, with allowance made for small gaps due to individual dead channels resulting from poor bond wire connections, physical sensor damage, or channels disabled during data acquisition. For each cluster, an effective segment number is calculated as the weighted average of all contributing segment numbers, using the energy deposit in each segment as the weighting factor.
    
    If any individual segment is empty, the entire clustering process is halted, as all segments must be populated in order to determine the earliest timestamp. The keepalive trigger events reduce the probability that any given segment will ever be empty; whenever a keepalive is sent along a specific elink, it is distributed to all segments managed by that link. Whenever a keepalive is selected to participate in a cluster, it is discarded, and the next earliest event in the same segment is instead evaluated.
    
    Due to the computationally-intensive nature of the cluster formation process, which requires examination of every segment to determine the earliest timestamp, a second algorithm was developed which implements further parallelism for subtasks within the clustering process. A schematic of the data flow through this parallel algorithm is shown in \figref{clustering}. One thread is dedicated to the first stage of clustering, accepting trigger events from all contributing elinks and sorting these events to the appropriate segments. In the second stage of clustering, multiple worker threads are used to form clusters in contiguous regions of the sensor, using the same algorithm as described in the serial model. However, when considering the parallel model, there is the possibility that a cluster may form which crosses the boundary between the regions covered by two different workers. If a cluster forms close enough to the edge of the region, it is tagged as possibly needing to be merged.
    
    \singleFig[!b]{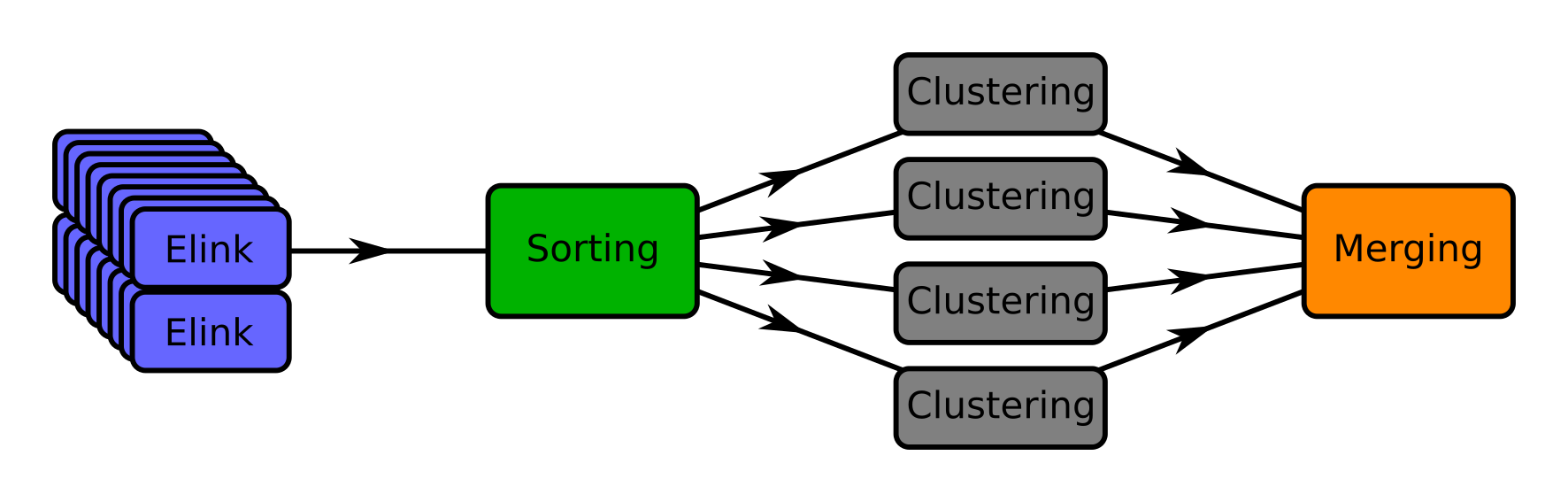}{clustering}{Schematic of data flow in the parallel clustering algorithm, which is a replacement for the serial grey element from \figref{pipeline}. Data from uplinks (blue) are first processed by a single thread (green), before being distributed to a pool of worker threads (grey), with each thread dedicated to a separate portion of the sensor area. Clusters formed by the worker threads are then merged by a single thread (orange) to form a single output data stream for the next stage of the pipeline. The remaining analysis then continues as shown in \figref{pipeline}.}{Schematic of data flow in the parallel clustering algorithm.}
    
    The third and final stage of the parallel clustering model is responsible for merging any clusters formed in the second stage. The algorithm is similar to that described for cluster formation within individual segments, although this thread now only considers forming larger clusters from two or more clusters which are tagged as candidates for merging. The result of this thread is a single datastream of time-ordered clusters, identical to the output of the serial model.
    
    \subsection{Hit Formation}
    \label{sec:online_hit}
    
    Cluster timestamps from both the p and n sides of the sensor are examined, and the earliest timestamp starts a coincidence window, as in the clustering process. All coincident clusters from both sides are brought into the same coincidence window, regardless of position, and all coincident combinations of one p-side and one n-side cluster are considered. Due to the angled p strips, not all combinations of clusters correspond to physical intersections within the sensitive area of the sensor, and candidate hits failing this criterion are rejected. For all other hits, the two-dimensional position on the sensor surface is calculated, and this position is used along with the equation of the sensor plane to calculate the three-dimensional lab frame position of the particle interaction.
    
    It is possible for the same cluster to contribute to the formation of more than one hit. In some cases, these hits represent random coincidences, in which two true interactions which happen to fall within the same coincidence window lead to three or four potential hit positions, producing `ghost' hits in addition to the true coincidences. In other cases, two coincident interactions which happen to trigger the same strips, or adjacent strips on one side of the sensor, may contribute to the same cluster, which would make duplicate use of the same cluster valid. No effort is made to reject ghost hits at this stage, other than the selection of a tight coincidence window facilitated by good timing resolution, as it is presumed that the majority of tracks containing these ghost hits will fail to pass the filters of fIVI, and it is more efficient to defer this processing to a less-busy thread.
    
    \subsection{Track Formation}
    \label{sec:online_track}
    
    The track formation process works very similarly to hit formation. Hit timestamps from sensors in the first and second layer are considered, forming a coincidence window in the same fashion. As with hit formation, no filters are applied during the track formation process, leading to the likelihood that false tracks will be formed. Such false tracks may be those containing a ghost hit, or those for which the coincidence between layers is only random. The rejection of these false tracks is deferred to the fIVI filters, which are applied during the vertex reconstruction process.
    
    \subsection{Vertex Reconstruction}
    \label{sec:online_vertex}
    
    The vertex formation process uses the fIVI algorithm, applying filters as previously described to reject candidate hits and tracks which are unlikely to correspond to fragments originating in the target region \autocite{hymers_intra-_2021}. Through this process, the majority of tracks formed from ghost hits are also rejected, as these tracks are less likely to form a trajectory which passes close to the target. The current vertex formation algorithm reconstructs each track independently, making this the only stage of reconstruction where no coincidence is required. This complete independence would allow a number of trivial parallel algorithms, such as a thread pool; however, for the limited number of sensors implemented in the fIVI Range Monitoring System, the data rate is not sufficient to fully utilize even a single core. Therefore, to reduce the overhead of inter-thread communication, a simple serial model is used for vertex reconstruction.
    
    \subsection{Range Monitoring}
    \label{sec:online_rm}
    
    As with vertex reconstruction, the range monitoring algorithm is an implementation of the algorithm previously described for use with fIVI. As vertices are reconstructed, they are projected along the nominal beam axis to produce a one-dimensional histogram. Once all vertices have been processed, the entire range shift determination algorithm is performed serially. First, the histogram is normalized such that its maximum value occurs at \qty{250}{\vertices\per\mm}. A linear search identifies the beginning of the distal edge, which is the feature used for range monitoring. A logistic fit to the distal edge is then performed, beginning at the identified start point and extending to a fixed maximum depth, using the Minuit/Migrad minimizer as implemented in ROOT \autocite{brun_root_1997}. A second logistic function is provided as input to the analysis software, describing a reference distal edge to which a range comparison is made. The two functions are translated vertically such that both lower asymptotes, beyond the distal edge, approach a limiting value of \qty{1}{\vertex\per\mm}. Then, the fit to the current dataset is translated horizontally to achieve the best agreement with the reference distal edge function, as determined by computing a discrete $\chi^2$ statistic between the two functions, evaluated every \qty{100}{\micro\m} within the fit range for the reference distal edge.
    
    Because the $\chi^2$ minimization is performed between two continuous functions, there is a single global minimum, and the $\chi^2$ statistic will continually decrease as the translation approaches the value which achieves minimization. Therefore, once an increase in the $\chi^2$ statistic is noted, it can be concluded that the minimum has been passed, and the search may be terminated early. This optimization reduces the number of range shifts which need to be evaluated, which is expected to reduce the time required to complete analysis.
    
    \section{Performance Measurement}
    \label{sec:online_performance}
    
    \subsection{Methods}
    \label{sec:online_methods}
    
    As the performance of this data acquisition system has been previously measured to accept data rates corresponding to clinical irradiation, current performance benchmarking was focused on the time needed to perform the analysis. To model online analysis, a driver program was used which replayed experimental data at the same rate it was originally collected. This driver program read an entire data set into memory, then sent the packets along a named pipe in the same fashion as would be employed for online analysis. To ensure that contention in the analysis software did not affect the playback rate, the single-threaded driver program was restricted to use a single physical CPU core, while the  analysis software was restricted to the other 23 physical cores of the i9 13900K processor used for testing. To dedicate maximum computing resources to analysis, and to avoid possible contention through sharing a core capable of running two concurrent logical threads, the core selected for the driver was one of the 16 efficiency cores.
    
    The data sets used in this playback were originally collected at the quality assurance beamline of the Heidelberg Ion-Beam Therapy Center (HIT, Heidelberg, Germany). Clinical beams from the HIT synchrotron were delivered into cylindrical poly-(methyl-methacrylate) (PMMA) phantoms, \qty{16}{\cm} and \qty{32}{\cm} in diameter, and \qty{15}{\cm} in height. The fIVI Range Monitoring System was placed with the front sensor \qty{4.5}{\cm} from the edge of each phantom, at a \ang{40} angle from the beam axis. Data were collected at a variety of beam energies and intensities for complete \qty{5}{\second} synchrotron spills, the maximum duration used in radiotherapy at HIT. The range monitoring performance of this data set is described in a separate publication \autocite{hymers_clinical_2025}.
    
    To assess the analysis performance of the system, timestamps were recorded for the sending of the first and last packets by the driver, along with endpoints such as the completion time of various pipeline stages. Monitoring was performed for only one endpoint at a time, with other endpoints disabled by the use of compiler flags, to prevent monitoring of one endpoint from affecting the time to reach a later endpoint. Each measurement was repeated 15 times, to account for variations in performance due to background tasks on the host system. The most important endpoint studied was the beam-off analysis time, representing the difference in time between the transmission of the last packet to the analysis pipeline, and the determination of a final range shift. This endpoint directly corresponded to the post-irradiation time required to determine whether an error in BP range had occurred. A number of variables were examined for their impact on these endpoints, including: the serial and parallel clustering algorithms; the number of worker threads used for each sensor side in the parallel clustering algorithm, from 1 to 32; the minimum sleep time while waiting on a SPSC queue, from \qtyrange{0}{100}{\micro\s}; the presence or absence of a sleep call while waiting on a SPSC queue; the input data rate, as affected by variations in treatment beam intensity and Bragg peak depth; and the duration of irradiation, from \qtyrange{0.5}{5}{\s}. As the experimental data set contained only complete spills, shorter durations were modelled by cutting off the data set at a GERI packet boundary partway through a spill. For all phases of analysis, a uniform \qty{31.25}{\nano\s} coincidence window is used, corresponding to five ticks of the timing clock.
    
    Performance-critical functions and blocks were also evaluated by the use of the `perf’ tool in Linux, and visualized by the `hotspot’ user interface. `perf’ operates on a hardware timer, pausing execution at a fixed frequency and evaluating the call stack for each thread at the time of the pause. Over a long execution time, this method provides an estimation of how much processing time is spent in each function, including calls to other functions. `hotspot’ generates plots from these data which show the aggregate layers of the call stack, with larger blocks representing functions which take up more processing time, and any time not covered by higher levels of the call stack representing computation which occurs directly within that function, rather than a function call. When analyzing threaded code, multiple threads which run the same process appear grouped within the same call stack, so normalization is required based on the number of threads assigned a particular task. Profiling using `perf' was completed for offline analysis, in which the entire data set was available for analysis immediately, to reduce the impact of waiting for the next data packet, and instead emphasize the most performance-sensitive functions in the analysis code.
    
    \subsection{Profiling}
    \label{sec:online_profiling}
    
    Profiling of offline analysis provides valuable insight into the most relative computational expense of various portions of the online analysis code. A visualization of the relative computational costs of the different stages is shown in \figref{perf_serial}. The most expensive portion of the pipeline is the cluster formation algorithm, using \qty{49.3}{\percent} of processing time, with only four threads. In contrast, the other analysis stages of the pipeline are responsible for between \qty{0.5}{\percent} and \qty{3.0}{\percent} of processing time, with one or two threads each. The other significant contributor to processing time is the individual processing of each elink, consuming \qty{35.5}{\percent} of processing time. However, as a separate thread is created for each elink, each individual thread is responsible for only approximately \qty{0.6}{\percent} of the total computation time.
    
    \rotFig{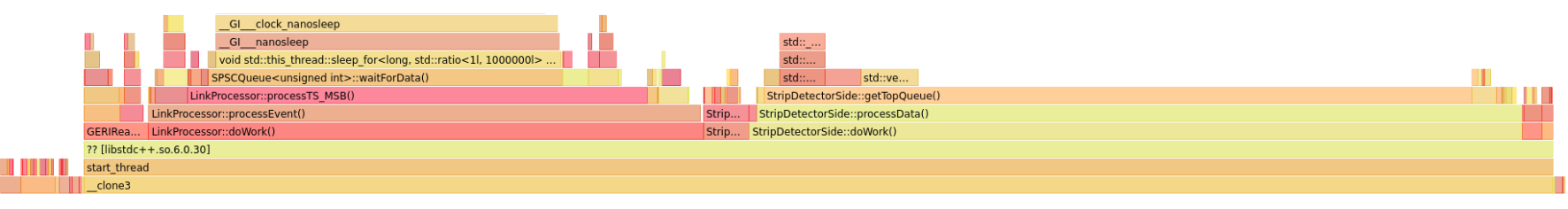}{perf_serial}{Performance visualization for analysis using a serial model for clustering. The width of each block is proportional to the fraction of compute time consumed by a given function, while the vertical stacking of blocks represents the function call hierarchy. The lowest few functions, which cover the majority of the horizontal axis, represent the system calls to create new threads. Due to the highly parallel nature of the analysis software, these calls underlie the majority of the compute time. No accounting is made in this visualization for the number of threads executing each stage of the pipeline.}{Performance visualization for analysis using a serial model for clustering.}
    
    The clustering process is clearly the most computationally-expensive stage of analysis. A naive division would suggest that each of the four sensor sides is responsible for \qty{12.3}{\percent} of the total computing resources, while the per-thread resource consumption of the other stages varies between \qty{0.5}{\percent} and \qty{1.5}{\percent}. However, it is more likely that the two threads responsible for the front sensor would require more computational resources than the two responsible for the rear sensor, due to the elevated event rate on the front sensor, which is placed closer to the target. Also of note is that, unlike the elink processing, in which more than half of the processing time is spent waiting on inter-thread communication, less than \qty{2}{\percent} of the resources consumed by clustering are spent waiting for other threads. The significantly elevated resource requirements of clustering relative to other pipeline stages, along with the rate-limiting behaviour indicated by a low fraction of inter-thread waiting time, support the decision to investigate a parallel version of the clustering process. Approximately \qty{70}{\percent} of the clustering workload is dedicated to identifying the segment with the earliest timestamp, indicating that optimization or parallelization of this routine would have the greatest impact on performance.
    
    Profiling a parallel version of the clustering process reveals a significant improvement in the balance between pipeline stages. The fraction of total processing time devoted to clustering remains high, at \qty{46}{\percent}, as shown in \figref{perf_mt}. However, this work is now spread out over more threads, bringing the maximum processing time time per thread as low as \qty{2}{\percent}, more in line with other stages of the pipeline. At four clustering workers per sensor side, the average per-thread workload of the sorting and clustering stages are similar, indicating that at more than four threads per sensor side, the bottleneck may transition from the clustering stage to the sorting stage. With such a transition, there may not be benefit to operating more than four workers per clustering stage. However, further performance improvements may be possible by also introducing parallelism in the sorting stage along with the clustering stage. The proportion of processing time dedicated to all other stages of the pipeline remains similar to the serial model.
    
    \rotFig{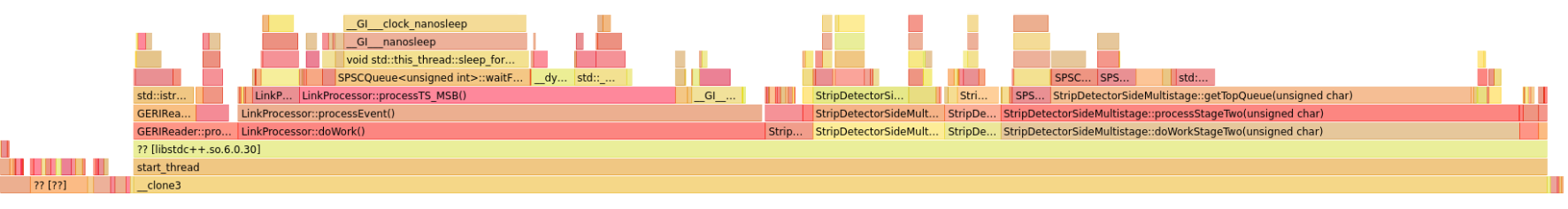}{perf_mt}{Performance visualization for analysis using a parallel model for clustering, with four stage-two worker threads per sensor side. The width of each block is proportional to the fraction of compute time consumed by a given function, while the vertical stacking of blocks represents the function call hierarchy. The lowest few functions, which cover the majority of the horizontal axis, represent the system calls to create new threads. Due to the highly parallel nature of the analysis software, these calls underlie the majority of the compute time. No accounting is made in this visualization for the number of threads executing each stage of the pipeline.}{Performance visualization for analysis using a parallel model for clustering.}
    
    \subsection{Worker Dependence}
    \label{sec:online_worker}
    
    The serial model of clustering demonstrates consistent performance, completing analysis in around \qty{250}{\milli\s} for most cases, although the lowest data rates exhibit improved performance. This model is strictly superior to a single worker and, for low data rates, is also superior to two worker threads per sensor side. The reduced performance of the parallel model with a low number of workers is attributed to the additional overhead of incorporating more inter-thread communication.
    
    With a low worker count, the parallel model demonstrates a longer beam-off analysis time for both larger data sets and higher data rates, with the largest data set and highest data rate requiring in excess of \qty{3.5}{\s} to process data from a \qty{5.0}{\s} spill. This relationship is believed to occur because data processing occurs more slowly than data acquisition, leading to a backlog of data to be processed in the beam-off time. This backlog is larger at higher data rates due to the higher rate of accumulation, and for longer irradiations due to the greater duration for which backlogged events are accumulated. The inter-run variation in beam-off analysis time is also much greater with only a single worker per sensor side, with a standard deviation as large as \qty{400}{\milli\s} for the highest-rate data sets, as shown in \figref{workers}.
    
    \singleFig[!b]{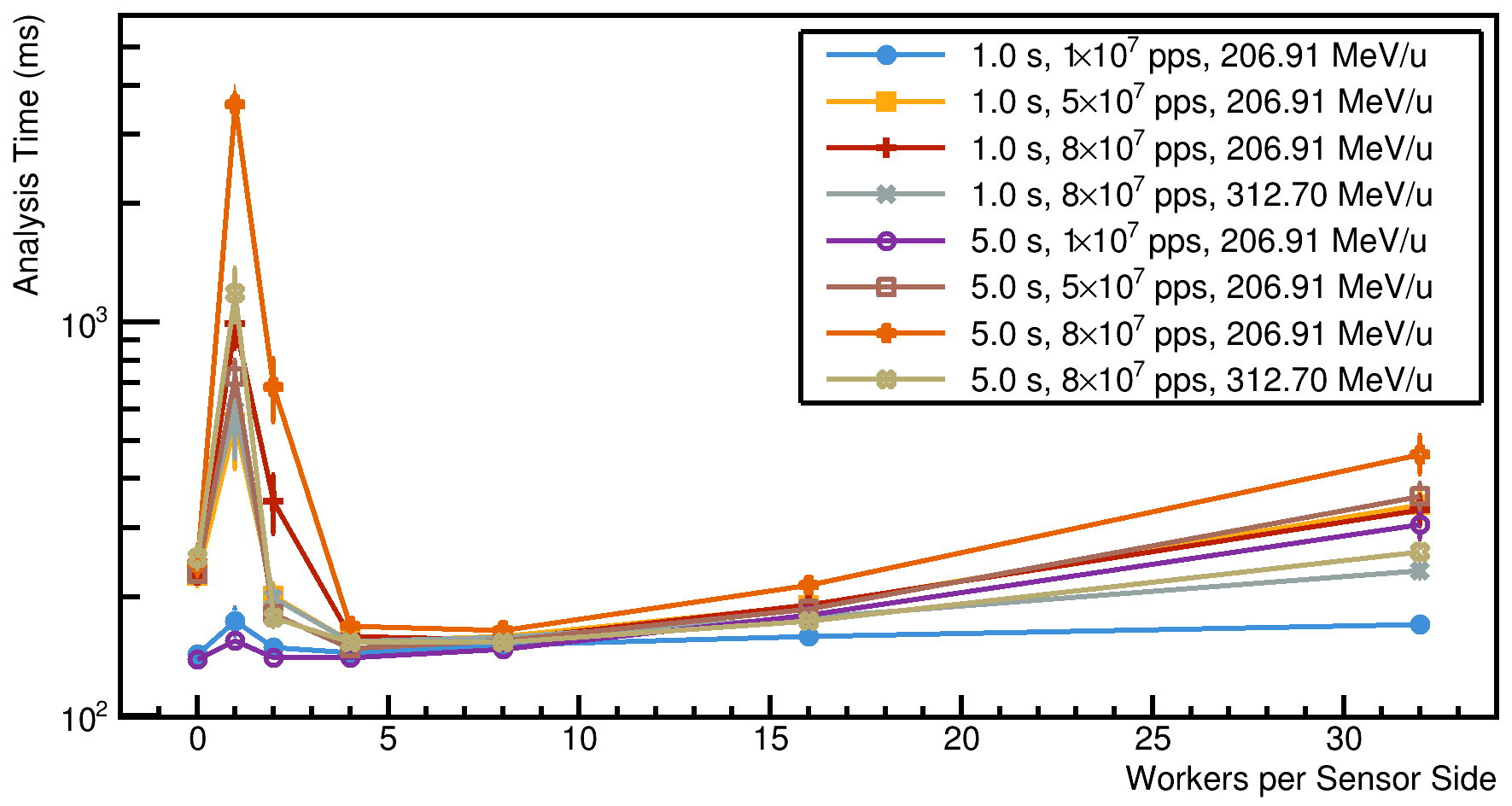}{workers}{Analysis time as a function of worker count per sensor side in the clustering phase. The data points at zero workers represent the serial model. The presented data is for a sleep time of \qty{30}{\micro\s}. A line has been drawn between points for each data series, to guide the eye. The parallel model yields reduced performance relative to the serial for low worker counts. At very high worker counts, contention effects reduce the advantage of parallelism, particularly at high data rates.}{Analysis time as a function of worker count per sensor side in the clustering phase.}
    
    As the number of workers increases, performance improves beyond the serial model, with all data sets converging to completion times between \qtyrange[range-phrase=~and~]{130}{170}{\milli\s} and inter-run standard deviations of less than \qty{7}{\milli\s} by four workers per side. Doubling the number of workers to eight per side does not meaningfully change these values, indicating that for this combination of hardware and data rates, between four and eight workers per sensor side is ideal.
    
    However, further increases to 16 or 32 clustering workers per sensor side result in increases to both the average completion time and inter-run standard deviation above the optimal values. Two factors are believed to contribute to this reversal. As the number of workers is increased, the proportion of clusters which will span between workers and must be merged by the single thread of the third stage increases. Therefore, the rate limitation shifts from the second to the third stage, and a further increase in the number of second-stage workers provides no benefit. Furthermore, as the number of workers increases, the total number of threads used by the analysis software likewise increases, leading to increased contention over the limited number of physical cores, and increasing the likelihood that a given thread will need to wait on a thread directly preceding or succeeding it in the pipeline. This increased contention is directly reflected in the worsened performance of the analysis software.
    
    \subsection{Rate Dependence}
    \label{sec:online_rate}
    
    \doubleFig[!b]{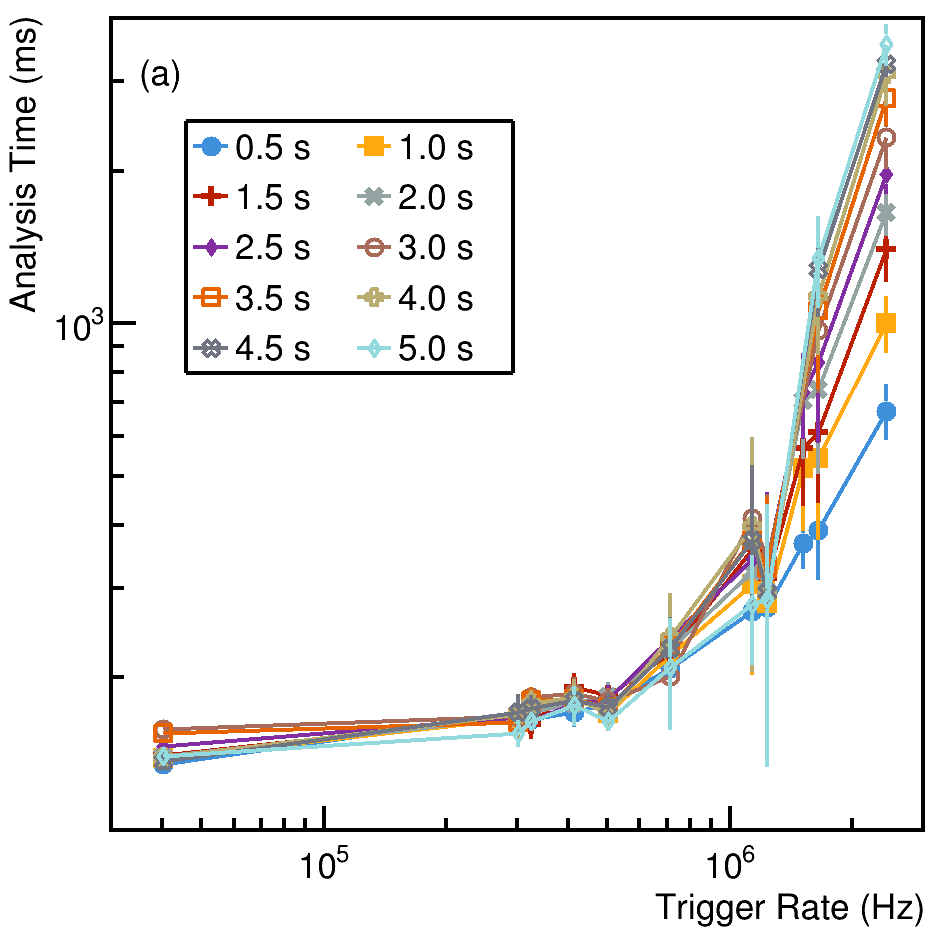}{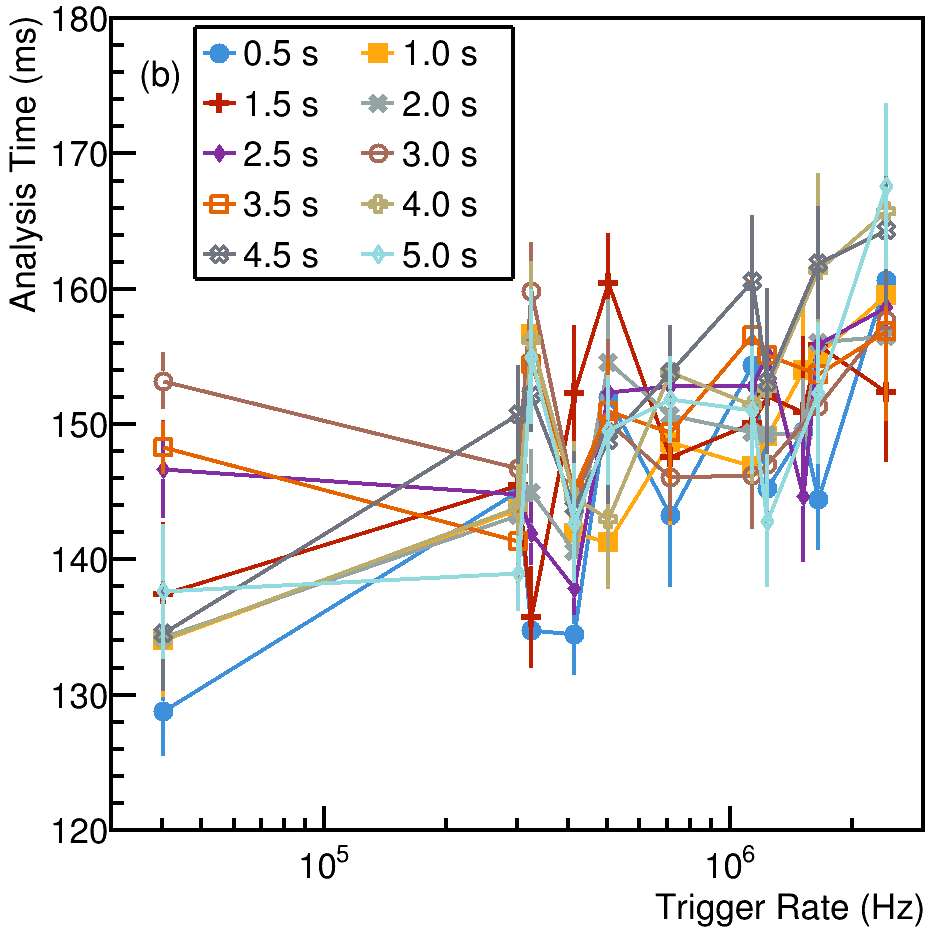}{rates}{Analysis time as a function of trigger rate through the data acquisition system. The presented data is for the parallel clustering model with (a) one worker per sensor side and (b) four workers per sensor side. The presented data uses a sleep time of \qty{30}{\micro\s}. A line has been drawn between points for each data series, to guide the eye. With sufficient worker count to avoid bottlenecks, there is a mild dependence of the total analysis time on the event rate, while at insufficient worker counts, this dependence is more pronounced. This dependence is explained by the plotted analysis time representing the time to both complete vertex formation in the last packet, and the time to perform the range shift analysis.}{Analysis time as a function of trigger rate through the data acquisition system.}
    
    When considering online analysis of data sets using the optimal four or eight clustering workers per sensor side, there is only a weak dependence of the analysis time on the data rate, as shown in \figref{rates}. This result is consistent with data from other endpoints, which indicates that each GERI packet is fully processed and the data is added to the histogram before the next packet is made available. There is an observable relationship between the incident event rate and the beam-off analysis time present at a variety of worker counts. At low worker counts, where there is insufficient time to process an entire high-data-rate packet before the next is delivered, the dominant contribution to this relationship represents the accumulation of a rate-dependent fraction of unprocessed events from each packet. However, the beam-off analysis time includes both the time needed to process the final packet and add the resultant vertices to the histogram, as well as to complete the range shift analysis. The processing time for the final packet clearly depends on the event rate, as is confirmed through analysis of endpoints other than beam-off analysis time. The extrapolation of this trend to an input rate of zero is consistent with the time requirement of the range shift determination algorithm, after the reconstruction of all interaction vertices.
    
    \subsection{Duration Dependence}
    \label{sec:online_dur}
    
    There is no significant dependence of either the beam-off analysis time or its standard deviation on the fraction of a complete spill which is delivered to the target, so long as the number of workers is sufficient. As with the rate dependence, this lack of correlation is reflective of each packet being fully processed and added to the histogram before the next packet has been made available. Furthermore, when the number of workers is insufficient, there is a linear dependence of the beam-off analysis time on the duration of the spill. This linear relationship, shown in \figref{duration}, reflects the additional processing time required for each packet beyond the \qty{100}{\milli\s} packet duration, accumulated over the total number of beam-on packets.
    
    \doubleFig[!b]{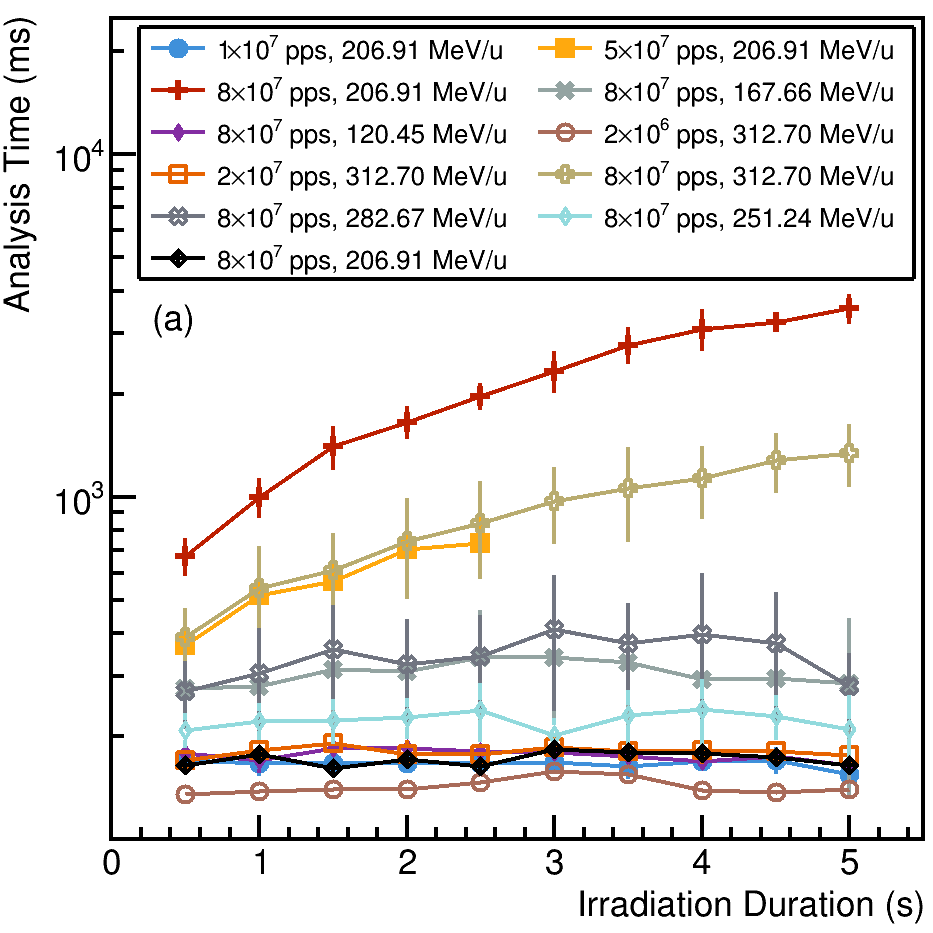}{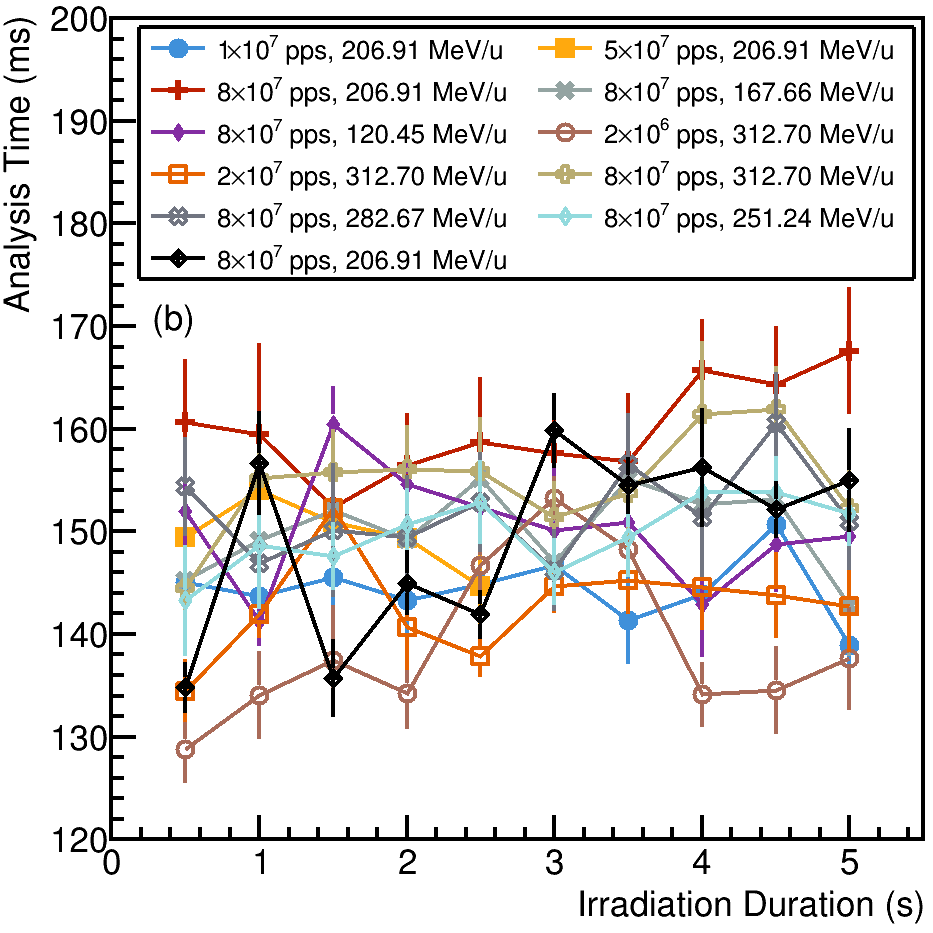}{duration}{Analysis time as a function of irradiation duration. The presented data is for the parallel clustering model with (a) one worker per sensor side and (b) four workers per sensor side. The presented data uses a sleep time of \qty{30}{\micro\s}. A line has been drawn between points for each data series, to guide the eye. With sufficient worker count to avoid bottlenecks, there is no clear relationship between the irradiation duration and the analysis time. However, at low worker count, there is a clear linear relationship in the high-rate data sets.}{Analysis time as a function of irradiation duration.}
    
    \subsection{Sleep Timing}
    \label{sec:online_sleep}
    
    The minimum sleep duration when one thread is waiting for another does not have any clear correlation with the analysis performance, so long as this duration is nonzero. When the sleep duration is zero, the waiting thread is not guaranteed to be descheduled, and may continue to consume CPU cycles which could otherwise be used to progress other tasks. This behaviour also caused playback to occur more slowly than expected when descheduling did not occur, due to contention between the playback and analysis executables causing background system tasks to run almost exclusively on the core used for playback. Due to the possibility of interfering with data acquisition, only a nonzero sleep duration should be considered for online analysis.
    
    \singleFig[!b]{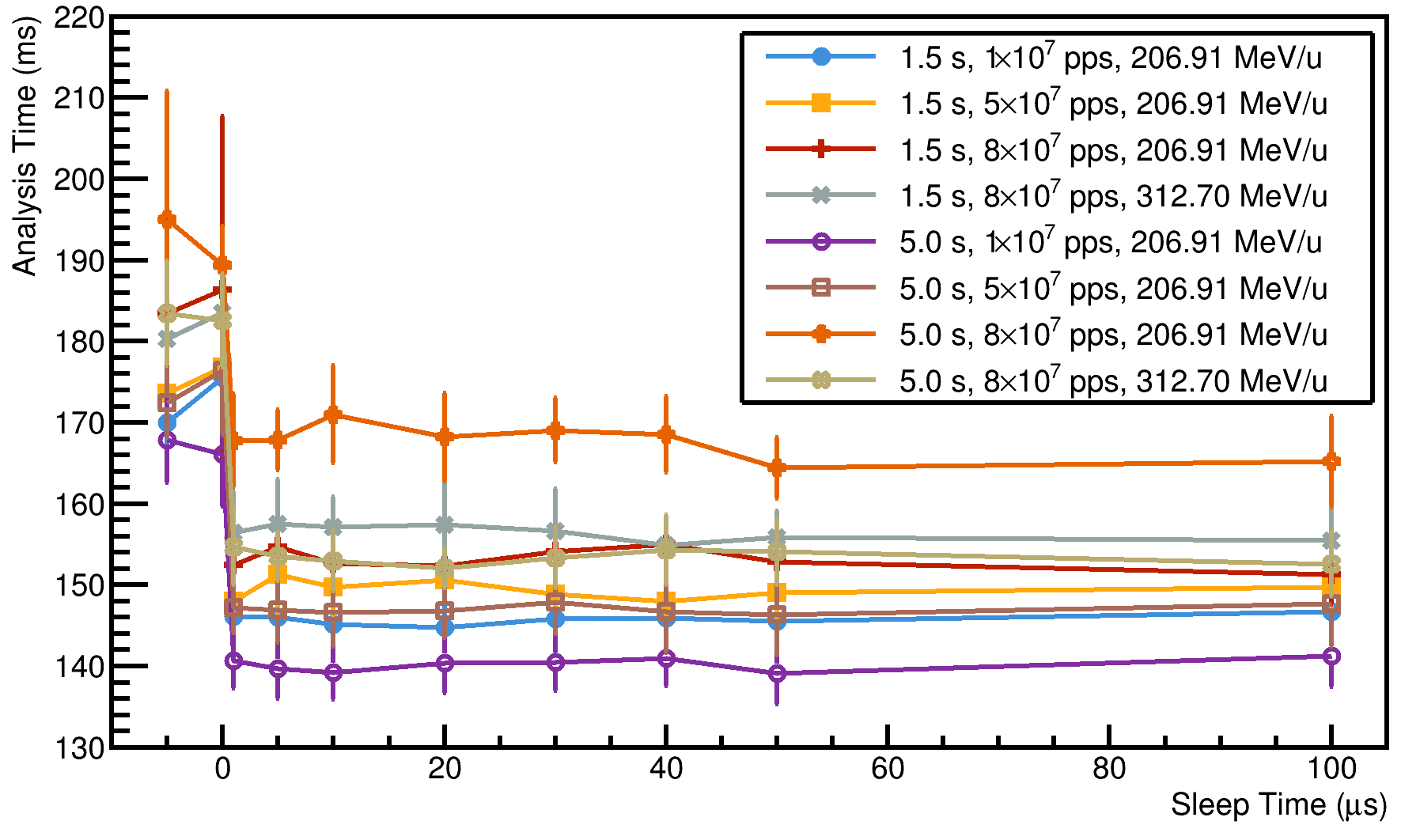}{sleep}{Analysis time as a function of sleep time when waiting for data from another thread. The presented data is for the parallel clustering model with four workers per sensor side. A line has been drawn between points for each data series, to guide the eye. The data point at negative sleep time corresponds to a complete removal of the sleep calls. There is no significant difference between any of the nonzero sleep durations tested. There is also no significant difference between a zero sleep duration and removing the sleep calls entirely. However, a significant difference does exist between zero and nonzero sleep durations, with a nonzero duration being associated with improved analysis performance.}{Analysis time as a function of sleep time when waiting for data from another thread.}
    
    As shown in \figref{sleep}, there is no significant performance difference observed between a sleep duration of zero and removing the sleep calls altogether, which supports the interpretation that descheduling does not necessarily occur when waiting for zero duration. Data collected during profiling also indicates that in both cases -- when sleep calls are removed or when the minimum sleep duration is zero -- the majority of processing at all stages is devoted to waiting on other threads, consuming over half of all processing time. The same behaviour occurs for all worker counts, indicating that within the tested bounds, an extended waiting time is unable to mitigate the effects of contention.
    
    While a longer sleep time might be expected to reduce contention, due to a reduction in the average number of active threads at any given time, this behaviour was not observed. Although it is possible that some additional pattern may emerge when considering a longer sleep time than the maximum tested duration of \qty{100}{\micro\s}, it is anticipated that the only significant performance impact in this case would be negative. As the wait time is extended, the probability that a thread will continue to sleep for a significant duration, even after being assigned work by a previous stage of the pipeline, is expected to increase.
    
    \section{Discussion}
    \label{sec:online_disc}
    
    For nearly all examined data sets, the clinical target for processing an entire \qty{5.0}{\s} spill in \qty{3.0}{\s} is met without requiring optimization. The sole exception is the data set with the highest data rate, which requires \qty{3.5}{\s} after beam-off for analysis to complete, when using a single clustering worker per sensor side. When running with an optimized four or eight workers per sensor side in the clustering phase, all data sets are capable of processing a partial or complete spill at any tested data rate within \qty{200}{\milli\s}, which is within the target for measurement during a spill pause, with or without reacceleration \autocite{schoemers_first_2017}. However, this duration is still in excess of the minimum quoted reacceleration time, indicating that further improvements in the speed of analysis could augment the clinical relevancy of range monitoring.
    
    The largest contribution to the beam-off analysis time is the range shift determination algorithm. In the tested software, this algorithm runs entirely sequentially; as the majority of the total CPU time is used in event building and reconstruction, that code has been the primary focus of optimization. However, opportunities exist to implement parallelism during the range shift determination algorithm as well. As this algorithm runs after the parallel portion of the analysis is complete, rather than during the pipeline, parallelism could be added without increasing contention. ROOT supports implicit multi-threading when performing fits, and while enabling this feature would be straightforward, the potential performance improvement is limited, as the range shift determination algorithm typically performs only a single fit \autocite{piparo_expressing_2017}. As the opportunity for implicit multi-threading is so limited, the performance penalty from spawning and communicating with extra threads may outweigh the benefit of parallelism. Another major opportunity for optimization in the range shift determination algorithm is the search for the translation of the current dataset which best matches the reference dataset. Implementing parallelism would allow multiple translations to be evaluated at once, using the existing algorithm. As this process is repeated for many independent configurations, the benefits of implementing a parallel algorithm are more likely to outweigh the overhead. However, given the relatively small size and short runtime of the rangefinding algorithm as compared to the remainder of the analysis, it will be necessary to carefully consider and benchmark the implementation of any form of parallelism, to determine whether it is beneficial or detrimental to overall performance. If the goal of online analysis is reframed from determining the exact range difference to assessing whether the range difference is within some margin of the expected value, as suggested by recent work \autocite{fischetti_inter-fractional_2020, kelleter_-vivo_2024}, an adapted algorithm might reduce the total number of evaluations which must occur, significantly reducing the required time.
    
    As the greatest impact on performance stems from selecting the proper number of worker threads for the clustering process, particular attention should be given to this parameter. While the tests performed in this work treated both sides of both sensors equally, the analysis software supports specifying individual worker counts for each sensor side. As the input data rates and output cluster rates are known to differ significantly when the sensors are placed \qty{12}{\cm} apart, additional benefit may be observed by tuning the number of workers to the individual sensor or sensor side. Optimizing the worker counts in this way would lead to a reduction in the total number of analysis threads, reducing contention while maintaining sufficient performance to process each packet before the next arrives. For sensors experiencing particularly low data rates, such as those placed at larger angles, or further away from the patient \autocite{finck_study_2017, ghesquiere-dierickx_investigation_2021}, contention could be further reduced by using the serial clustering model, rather than a low worker count. The reduction in contention afforded by these optimizations is expected to facilitate extension of the current high performance to faster input count rates.
    
    A further possible optimization in worker count could be achieved by allowing variation in the number of segments supported by each clustering worker. As the data produced by clinical irradiation has a strong dependence on angle, the workers managing cluster formation in more active regions of the sensor are naturally busier than those managing less busy regions. As the number of worker threads already exceeds the number of physical cores on the analysis system, reducing the number of threads, and therefore the load on the scheduler, could lead to meaningful performance improvements. However, allowing the transition between second-stage workers to occur at arbitrary segments could have an unanticipated impact on the performance of the third stage of clustering. In the tested configuration, for worker counts up to eight per sensor side, all transitions between second-stage workers occur at ASIC boundaries. Due to the use of individual analog microcables for each ASIC, and the reduced shielding at the edges of each microcable, these ASIC boundary segments are more sensitive to noise, and typically have higher thresholds to compensate \autocite{rodriguez_rodriguez_advancements_2025}. Therefore, these segments have reduced sensitivity, and the probability of a cluster spanning two or more workers is suppressed, reducing the workload on the third stage of clustering below what would be expected if all segments were equally sensitive. Allowing the transition between second-stage workers to occur at arbitrary channels should therefore be tested carefully to avoid creating a further bottleneck at the third stage of clustering.
    
    A significant number of the total threads created in the analysis process, as well as total computational resources, are those dedicated to trigger event formation, where one thread is currently used to process each uplink independently. As these threads not only process the trigger event data, but also keep track of timing for each link through the TS\_MSB frames, they represent a significant fraction of the resource utilization of the analysis software. However, the operations these threads perform are quite simple, consisting primarily of unpacking data and counting timestamp frames. In principle, these operations could be implemented in hardware, either at the level of the GBTxEMU board or the GERI, with a small latency cost. Indeed, the possibility of this processing occurring in hardware was considered during design of the communication protocol \autocite{kasinski_protocol_2016}. Performing this computation in hardware would also greatly reduce the data rate entering the data acquisition PC, as the constant overhead of timing data on each link would no longer be required. As can be seen in \figref{perf_serial} and \figref{perf_mt}, removing this task from the CPU would also lead to a significant reduction in the total computational power required.
    
    Another hardware optimization, which could be performed either in conjunction or independently, would be the delivery of separate data streams through the GERI. In the current system, each uplink produces a separate stream until being processed by the GERI, at which point the data streams are combined \autocite{zabolotny_versatile_2023}. However, the first step of the analysis software is to separate these streams again, which must be done serially, and is a potential bottleneck at higher data rates. At the minor cost of maintaining a larger number of file handles, separating the data stream for each link could potentially allow much higher performance, by eliminating a potential bottleneck at the earliest stage of processing.
    
    \section{Conclusion}
    \label{sec:online_conclusion}
    
    When operated with parallel cluster formation, the online fIVI analysis software demonstrates for the first time the consistent ability to determine a shift in BP depth in less than \qty{200}{\milli\s}. This performance will allow implementation of quality assurance at all energy changes between spills. For treatment systems which allow spill pauses, with or without energy changes, range shift determination may also be performed during each spill pause, with only a minor constraint on pause duration. Scaling to higher beam intensities or data rates is supported by the parallel structure of data analysis; this scaling could be further supported by moving the timestamp processing into hardware. Incorporating online monitoring of range shifts into clinical treatment would allow detection of range errors caused by anatomical or patient positioning changes between fractions, and provides the opportunity for treatment to be paused or aborted before completion if a significant range difference is detected. Through the combination of online monitoring with a precise implementation of IVI, a reduction in safety margins for fIVI-monitored treatment may also be possible, allowing a reduction in the dose to healthy tissue even for correctly-delivered fractions. The demonstration of rapid online feedback regarding ion range therefore represents a powerful tool for improving the consistency and safety of carbon ion radiotherapy.

    % References
    \printbibliography[title=References]

\end{document}